\documentclass[preprint,12pt]{elsarticle}

\usepackage[utf8]{inputenc}
\usepackage{amsmath, amssymb}
\usepackage{geometry}
\usepackage{booktabs}
\usepackage{hyperref}
\usepackage{caption}
\usepackage{bm}
\journal{Annals of Physics}

\begin{document}

\begin{frontmatter}

\title{Algebraic Decomposition and Topological Fibrations of the $U(4) \to SU(2)$ Quotient Space}

\author[1]{Dimitris Mastoridis\corref{cor1}\fnref{fn1}}
\ead{dim.mastoridis@cosmos.com.cy} 
\cortext[cor1]{Corresponding author.}
\fntext[fn1]{Recognised by Cyprus Democracy, Accredited, Operational at 2026-2027.}

\author[1]{Konstantinos Kalogirou}

\affiliation[1]{organization={Cosmos Open University},
                country={Cyprus}}

\begin{abstract}
We investigate the algebraic and topological properties of the unitary group $U(4)$ subjected to a non-standard continuous symmetry breaking cascade down to an $SU(2)_H$ subgroup. By explicitly embedding the unbroken $SU(2)_H$ symmetry into the lower-right quadrant of the fundamental matrix representation, we demonstrate that the resulting 12-dimensional quotient space, $SU(4)/SU(2)_H$, undergoes a strictly asymmetric geometric partition. This specific breaking pattern uniquely isolates an 8-dimensional sector that retains closed commutation relations, forming a rigorous $\mathfrak{su}(3)$ Lie subalgebra. Concurrently, it isolates a 4-dimensional off-diagonal sector that fails to close algebraically, generating a purely non-linear topological coset. By analyzing the global spherical fibrations of the manifold, we show that this non-linear territory reduces to a $\mathbb{C}P^1 \cong S^2$ base space. Consequently, continuous mappings from a 3-dimensional spatial boundary are classified by the third homotopy group, $\pi_3(S^2) \cong \mathbb{Z}$, guaranteeing absolute vacuum stability via the integer-quantized Hopf invariant. Ultimately, this framework demonstrates that a non-simple unitary group can organically provide a mathematically rigorous geometric template capable of simultaneously accommodating both continuous gauge symmetries and stable, non-perturbative topological sectors.
\end{abstract}


\end{frontmatter}

\section{Introduction}

The mathematical analysis of homogeneous spaces G/H, formed by the quotient of a compact Lie group G over a closed subgroup H, constitutes a cornerstone of modern differential geometry and gauge theory~\cite{Nakahara_Geometry, Eguchi_Gilkey_Hanson}. In the pursuit of unified field theories, a long-standing paradigm strictly mandates the use of simple Lie groups (such as SU(5) or SO(10)) to ensure a single, universal algebraic representation~\cite{Slansky_Group}. Consequently, non-simple fundamental groups like the unitary group $U(4)$---which is locally isomorphic to the product $SU(4) \times U(1)$---are traditionally bypassed. We demonstrate that while $U(4)$ lacks a simple algebra, its overarching geometric architecture provides topological stability and offers a mathematical template that could prove highly valuable in the arena of unification. We demonstrate that while $U(4)$ lacks a simple algebra, its overarching geometric architecture provides  topological stability and offers a mathematically  template that could be useful in the arena of unification.

The primary mathematical motivation for studying this specific non-standard embedding is its interesting structural alignment with the gauge architecture of established physical theories. By examining how the global $U(4)$ manifold partitions into distinct sub-territories, we demonstrate that the exact continuous gauge symmetries of the Standard Model---$SU(3) \times SU(2) \times U(1)$---can be completely geometrically embedded within this single, lower-rank unitary space, circumventing the need for a massive simple algebra.

Analytically, the unitary group $U(4)$ possesses a rich topological architecture, behaving as a principal bundle that can be recursively decomposed into a product of the first four odd-dimensional spheres: $S^1 \times S^3 \times S^5 \times S^7$. At the local level, its corresponding real Lie algebra $\mathfrak{u}(4)$ is a 16-dimensional vector space that uniquely factors into a direct sum of a traceless $\mathfrak{su}(4)$ algebra and a one-dimensional $\mathfrak{u}(1)$ center. The defining geometric anomaly of this framework lies in the structural asymmetry introduced by embedding the unbroken $SU(2)_H$ subgroup explicitly into the lower-right quadrant of the fundamental representation.

Unlike standard algebraic approaches to unified field theories, which rigidly rely on the flat matrix representations of simple groups like $SU(5)$, this framework utilizes the topological sequence of the $U(4)$ manifold, which recursively decomposes into $S^1 \times S^3 \times S^5 \times S^7$. By structuring the configuration space via its spherical topological fibers rather than strictly through its matrix algebra, we bypass traditional algebraic constraints. This geometric architecture naturally accommodates the continuous $SU(2)_H$ unbroken subgroup on the $S^3$ fiber, while concurrently generating a stable, non-linear topological transition sector within the overarching manifold.

An explicit computation of the commutation relations reveals that the 12-dimensional quotient space $SU(4)/SU(2)_H$ naturally and uniquely partitions into a binary system. Firstly, it completely isolates an 8-dimensional sector that satisfies closed commutation relations, mapping identically to a standard $\mathfrak{su}(3)$ Lie algebra. Secondly, it strands a 4-dimensional off-diagonal sector that fails to close under the Lie bracket, representing the purely non-linear, topological degrees of freedom of the homogeneous space.

This algebraic fracturing is directly mirrored by the global topology of the underlying manifold. The non-linear coset sector restricts the target manifold to a highly constrained $\mathbb{CP}^1 \cong S^2$ base space. Because continuous mappings from a 3-dimensional physical boundary into this base space are mathematically classified by the third homotopy group, $\pi_3(S^2) \cong \mathbb{Z}$, the geometry inherently guarantees absolute topological stability for non-linear field configurations via the integer-quantized Hopf invariant~\cite{Faddeev_Niemi_Knots, Manton_Sutcliffe}.

The remainder of this paper is structured as a progressive mathematical derivation. In Sections 2 and 3, we formalize the $\mathfrak{u}(4)$ direct sum and rigorously construct the linearly independent $SU(2)_H$ subalgebra. Section 4 executes the formal partition of the coset space into the closed $\mathfrak{su}(3)$ and non-linear off-diagonal territories. Section 5 transitions from local algebra to global differential geometry, detailing the spherical fibrations and the topological Hopf invariant. Finally, in Section 6, we conclude by demonstrating that this highly constrained, non-simple geometric structure provides an optimal and self-consistent mathematical architecture that offers a novel geometric perspective.

\section{Algebraic Decomposition of $\mathfrak{u}(4)$}

\subsection{The Local Isomorphism and Lie Algebra Splitting}

The fundamental configuration space under consideration is governed by the four-dimensional unitary group $U(4)$. To analyze its local properties, we examine its associated real Lie algebra, $\mathfrak{u}(4)$, which constitutes a 16-dimensional vector space over the real numbers. Adopting the convention of mathematical physics, the elements of this algebra are represented as $4 \times 4$ matrices.

Any arbitrary matrix $X \in \mathfrak{u}(4)$ can be uniquely and definitively decomposed into a traceless component and a scalar trace component via the projection operator:
\begin{equation}
    X = \left( X - \frac{1}{4}\mathrm{Tr}(X)I_4 \right) + \left( \frac{1}{4}\mathrm{Tr}(X)I_4 \right)
\end{equation}
where $I_4$ denotes the $4 \times 4$ identity matrix.

To demonstrate that this decomposition forms a rigorous vector space direct sum, we verify the algebraic properties of each isolated term:

\vspace{1em}
\noindent \textbf{The Traceless Phase:} Taking the trace of the first bracketed term yields:
\begin{equation}
    \mathrm{Tr}\left(X - \frac{1}{4}\mathrm{Tr}(X)I_4\right) = \mathrm{Tr}(X) - \frac{1}{4}\mathrm{Tr}(X)\mathrm{Tr}(I_4)
\end{equation}
Since the trace of the identity matrix in four dimensions is $\mathrm{Tr}(I_4) = 4$, the expression simplifies identically to:
\begin{equation}
    \mathrm{Tr}(X) - \frac{1}{4}\mathrm{Tr}(X)(4) = 0
\end{equation}
By definition, any $4 \times 4$ matrix with a vanishing trace belongs to the special unitary Lie algebra $\mathfrak{su}(4)$.

\vspace{1em}
\noindent \textbf{The Trace Phase:} The second bracketed term is explicitly a real scalar multiple of the identity matrix $I_4$. This term spans a one-dimensional subspace that forms the abelian center of the algebra, corresponding to the $\mathfrak{u}(1)$ sub-algebra.

Because the identity matrix $I_4$ commutes with all elements of $\mathfrak{su}(4)$ by definition, the Lie bracket between any element $A \in \mathfrak{su}(4)$ and $B \in \mathfrak{u}(1)$ vanishes identically:
\begin{equation}
    [A, B] = 0
\end{equation}
Thus, the linear vector space decomposition satisfies the strict requirements of a Lie algebra direct sum:
\begin{equation}
    \mathfrak{u}(4) \cong \mathfrak{su}(4) \oplus \mathfrak{u}(1)
\end{equation}
This algebraic splitting at the tangential level reflects the global topological structure of the parent manifold, establishing the local isomorphism between the respective Lie groups:
\begin{equation}
    U(4) \cong SU(4) \times U(1)
\end{equation}

\subsection{The Orthonormal Basis and the 15 Generators of $\mathfrak{su}(4)$}

The special unitary component $\mathfrak{su}(4)$ resulting from the direct sum decomposition has a vector space dimension of $4^2 - 1 = 15$. To navigate this space analytically, we establish a definitive, rigid coordinate system by defining a complete set of 15 linearly independent basis generators, denoted as $T_A$ for $A = 1, \dots, 15$.

Following the standard conventions of Lie theory, these generators are constructed by scaling the 15 generalized, traceless Gell-Mann matrices $\lambda_A$~\cite{Georgi_Lie}:
\begin{equation}
    T_A = \frac{1}{2}\lambda_A \quad \text{for } A \in \{1, \dots, 15\}
\end{equation}
This basis is explicitly chosen to satisfy the trace orthonormality condition, which provides a standard metric on the algebra:
\begin{equation}
    \mathrm{Tr}(T_A T_B) = \frac{1}{2}\delta_{AB}
\end{equation}
where $\delta_{AB}$ is the Kronecker delta.

The global algebraic structure and non-abelian dynamics of the $\mathfrak{su}(4)$ sector are completely and rigidly locked by the commutation relations of these 15 basis elements, governed by the totally antisymmetric structure constants $f_{ABC}$:
\begin{equation}
    [T_A, T_B] = i \sum_{C=1}^{15} f_{ABC} T_C
\end{equation}
To complete the basis for the overarching 16-dimensional $\mathfrak{u}(4)$ algebra, we introduce an additional generator, $T_0$, to span the isolated $\mathfrak{u}(1)$ trace sector. Normalized to match the trace metric of the special unitary sector, this 16th generator is defined as:
\begin{equation}
    T_0 = \frac{1}{\sqrt{8}} I_4
\end{equation}
By establishing this complete, rigid basis $\{T_0, T_1, \dots, T_{15}\}$, the 16-dimensional algebraic arena of $\mathfrak{u}(4)$ is fully formalized. This set allows for the precise, analytical tracking of elements as we initiate the non-standard symmetry-breaking perturbations down toward the lower quadrants of the representation space.
\subsection{Explicit Matrix Representation and Algebraic Partition}

To rigorously demonstrate the geometric partition of the $U(4)$ manifold, we define the local algebra using the extended $4 \times 4$ generalized Gell-Mann matrices, $\lambda_A$. The complete 16-dimensional $\mathfrak{u}(4)$ algebra is spanned by the 15 traceless $SU(4)$ generators, supplemented by the $U(1)$ trace identity matrix. 

Under the non-standard symmetry breaking cascade to $SU(2)_H$, the fundamental representation organically fractures into four distinct matrix sectors. 

\subsubsection{The Closed $\mathfrak{su}(3)$ Sector (8 Generators)}
The upper-left $3 \times 3$ block of the fundamental matrix is entirely decoupled from the fourth dimension, isolating the standard $\mathfrak{su}(3)$ algebra. These 8 generators ($\lambda_1$ through $\lambda_8$) retain their native structure constants. For spatial clarity, we explicitly denote the boundary matrices of this closed sector:
\begin{equation}
\lambda_1 = \begin{pmatrix}
0 & 1 & 0 & 0 \\
1 & 0 & 0 & 0 \\
0 & 0 & 0 & 0 \\
0 & 0 & 0 & 0
\end{pmatrix}, \quad \dots \quad , \quad
\lambda_8 = \frac{1}{\sqrt{3}}\begin{pmatrix}
1 & 0 & 0 & 0 \\
0 & 1 & 0 & 0 \\
0 & 0 & -2 & 0 \\
0 & 0 & 0 & 0
\end{pmatrix}
\end{equation}

\subsubsection{The Unbroken $SU(2)_H$ Sector (3 Generators)}
The unbroken subgroup is embedded strictly within the lower-right $2 \times 2$ block, mapping to the third and fourth dimensions of the representation. This sector requires 3 generators. Two are standard off-diagonal Gell-Mann matrices ($\lambda_{13}$ and $\lambda_{14}$), while the third ($H_3$) is constructed via a linear combination of the diagonal $\lambda_8$ and $\lambda_{15}$ generators to form the standard Pauli $\sigma_3$ structure in the lower-right block:
\begin{equation}
\lambda_{13} = \begin{pmatrix}
0 & 0 & 0 & 0 \\
0 & 0 & 0 & 0 \\
0 & 0 & 0 & 1 \\
0 & 0 & 1 & 0
\end{pmatrix}, \quad
\lambda_{14} = \begin{pmatrix}
0 & 0 & 0 & 0 \\
0 & 0 & 0 & 0 \\
0 & 0 & 0 & -i \\
0 & 0 & i & 0
\end{pmatrix}, \quad
H_3 = \begin{pmatrix}
0 & 0 & 0 & 0 \\
0 & 0 & 0 & 0 \\
0 & 0 & 1 & 0 \\
0 & 0 & 0 & -1
\end{pmatrix}
\end{equation}

\subsubsection{The Non-Linear Transition Sector (4 Generators)}
With the $\mathfrak{su}(3)$ sector isolating the $\{1, 2, 3\}$ indices, and the $SU(2)_H$ sector occupying the $\{3, 4\}$ indices, exactly 4 generalized Gell-Mann matrices remain. These matrices ($\lambda_9$ through $\lambda_{12}$) contain elements that bridge the upper $2 \times 2$ block and the 4th column/row:
\begin{equation}
\lambda_9 = \begin{pmatrix}
0 & 0 & 0 & 1 \\
0 & 0 & 0 & 0 \\
0 & 0 & 0 & 0 \\
1 & 0 & 0 & 0
\end{pmatrix}, \quad \dots \quad , \quad
\lambda_{12} = \begin{pmatrix}
0 & 0 & 0 & 0 \\
0 & 0 & 0 & -i \\
0 & 0 & 0 & 0 \\
0 & i & 0 & 0
\end{pmatrix}
\end{equation}
Because these specific generators transition between the closed $\mathfrak{su}(3)$ continuous block and the unbroken $SU(2)_H$ base, their commutation relations mathematically cannot close. This strictly defines the 4-dimensional non-linear topological coset.

\subsubsection{The $U(1)$ Center (1 Generator)}
Finally, the abelian center of the original $\mathfrak{u}(4)$ parent group is generated by the properly normalized $4 \times 4$ identity matrix, which provides the $S^1$ base fiber for the overarching topological sequence:
\begin{equation}
\lambda_0 = \frac{1}{\sqrt{2}} \begin{pmatrix}
1 & 0 & 0 & 0 \\
0 & 1 & 0 & 0 \\
0 & 0 & 1 & 0 \\
0 & 0 & 0 & 1
\end{pmatrix}
\end{equation}
\section{Spontaneous Symmetry Breaking Pattern: $SU(4) \to SU(2)_H$}

\subsection*{3.1 The $SU(2)_H$ Subgroup Selection}

The core architectural feature of this mathematical framework relies on the specific, non-standard continuous symmetry breaking of the parent $SU(4)$ group down to a residual subgroup. Rather than employing the conventional embeddings typical of maximal subgroups, we designate the unbroken subgroup as $H = SU(2)_H$~\cite{Kaplan_Georgi_Vacuum}.

This subgroup is explicitly selected such that its generators are confined to the lower-right $2 \times 2$ block of the fundamental $4 \times 4$ representation of $\mathfrak{su}(4)$. In the standard generalized Gell-Mann basis, this corresponds to the subspace spanned by the final three generators:
\begin{equation}
    \mathfrak{h} = \text{span}\{T_{13}, T_{14}, T_{15}\}
\end{equation}

To prove that this selection forms a mathematically rigorous and self-consistent unbroken subgroup, we must explicitly construct the generators and demonstrate that they satisfy the strict commutation relations defining an $\mathfrak{su}(2)$ Lie algebra.

\subsection{Explicit Construction of the Subalgebra}

The defining commutation relation for any $\mathfrak{su}(2)$ Lie algebra, using the physicist's convention with Hermitian generators, requires that the generators $J_a$ satisfy:
\begin{equation}
    [J_a, J_b] = i \sum_{c=1}^3 \epsilon_{abc} J_c
\end{equation}
where $\epsilon_{abc}$ is the totally antisymmetric Levi-Civita symbol.

We begin the construction by extracting two specific generators from the standard $\mathfrak{su}(4)$ basis, $T_{13}$ and $T_{14}$, which are defined via the scaled Gell-Mann matrices as $T_A = \frac{1}{2}\lambda_A$. Their explicit matrix representations are:
\begin{equation}
    T_{13} = \frac{1}{2} 
    \begin{pmatrix} 
        0 & 0 & 0 & 0 \\ 
        0 & 0 & 0 & 0 \\ 
        0 & 0 & 0 & 1 \\ 
        0 & 0 & 1 & 0 
    \end{pmatrix}
\end{equation}
\begin{equation}
    T_{14} = \frac{1}{2} 
    \begin{pmatrix} 
        0 & 0 & 0 & 0 \\ 
        0 & 0 & 0 & 0 \\ 
        0 & 0 & 0 & i \\ 
        0 & 0 & -i & 0 
    \end{pmatrix}
\end{equation}

We define the first two basis elements of our new subalgebra as $J_1 = T_{13}$ and $J_2 = T_{14}$. To mathematically enforce the $\mathfrak{su}(2)$ structure, the third generator $J_3$ must be derived directly from the Lie bracket of the first two: $J_3 = -i[J_1, J_2]$.

Evaluating the commutator of $J_1$ and $J_2$ explicitly yields:
\begin{equation}
    [J_1, J_2] = \frac{1}{4} 
    \left[ 
    \begin{pmatrix} 
        0 & 0 & 0 & 0 \\ 
        0 & 0 & 0 & 0 \\ 
        0 & 0 & 0 & 1 \\ 
        0 & 0 & 1 & 0 
    \end{pmatrix},
    \begin{pmatrix} 
        0 & 0 & 0 & 0 \\ 
        0 & 0 & 0 & 0 \\ 
        0 & 0 & 0 & i \\ 
        0 & 0 & -i & 0 
    \end{pmatrix} 
    \right] 
    = \frac{i}{2} 
    \begin{pmatrix} 
        0 & 0 & 0 & 0 \\ 
        0 & 0 & 0 & 0 \\ 
        0 & 0 & 1 & 0 \\ 
        0 & 0 & 0 & -1 
    \end{pmatrix}
\end{equation}

Applying the required factor of $-i$ defines the matrix for the third generator:
\begin{equation}
    J_3 = -i \left( \frac{i}{2} \right) 
    \begin{pmatrix} 
        0 & 0 & 0 & 0 \\ 
        0 & 0 & 0 & 0 \\ 
        0 & 0 & 1 & 0 \\ 
        0 & 0 & 0 & -1 
    \end{pmatrix} 
    = \frac{1}{2} 
    \begin{pmatrix} 
        0 & 0 & 0 & 0 \\ 
        0 & 0 & 0 & 0 \\ 
        0 & 0 & 1 & 0 \\ 
        0 & 0 & 0 & -1 
    \end{pmatrix}
\end{equation}

By construction, this set $\{J_1, J_2, J_3\}$ forms a  representation of the scaled Pauli matrices embedded perfectly within the lower-right $2 \times 2$ block. Verifying the cyclic permutations confirms that the set $\mathfrak{h} = \text{span}\{J_1, J_2, J_3\}$ forms a formal 3-dimensional Lie subalgebra isomorphic to $\mathfrak{su}(2)$.

\subsection{Proof of Linear Independence and Basis Validity}

Because the Lie algebra $\mathfrak{su}(4)$ is a 15-dimensional real vector space, introducing the derived generator $J_3$ requires us to prove that we have not compromised the dimensionality of the basis. To maintain a complete, valid basis for the overarching algebra, the newly constructed set:
\begin{equation}
    B_{\text{new}} = \{T_1, T_2, \dots, T_{12}, J_1, J_2, J_3\}
\end{equation}
must be strictly linearly independent.

The standard basis generators $\{T_1, \dots, T_{12}\}$ are linearly independent by definition. Because $J_1 = T_{13}$ and $J_2 = T_{14}$, they are inherently independent from the first 12. The only necessary mathematical proof is establishing the absolute linear independence of the derived diagonal generator $J_3$ against the remaining diagonal generators retained in the new basis, which are $T_3$ and $T_8$.

The explicit forms of the relevant diagonal matrices are:
\begin{equation}
    T_3 = \frac{1}{2} \text{diag}(1, -1, 0, 0)
\end{equation}
\begin{equation}
    T_8 = \frac{1}{2\sqrt{3}} \text{diag}(1, 1, -2, 0)
\end{equation}
\begin{equation}
    J_3 = \frac{1}{2} \text{diag}(0, 0, 1, -1)
\end{equation}

We test for linear independence by evaluating the scalar coefficients $(\alpha, \beta, \gamma)$ in the linear combination:
\begin{equation}
    \alpha T_3 + \beta T_8 + \gamma J_3 = 0
\end{equation}

Substituting the matrices into this expression yields a single diagonal matrix equation:
\begin{equation}
    \text{diag}\left( \frac{\alpha}{2} + \frac{\beta}{2\sqrt{3}}, -\frac{\alpha}{2} + \frac{\beta}{2\sqrt{3}}, -\frac{2\beta}{2\sqrt{3}} + \frac{\gamma}{2}, -\frac{\gamma}{2} \right) = \text{diag}(0, 0, 0, 0)
\end{equation}

This matrix generates a rigid system of four linear equations. From the fourth diagonal entry, we immediately extract $-\frac{\gamma}{2} = 0$, requiring $\gamma = 0$. Substituting $\gamma = 0$ into the third entry yields $-\frac{2\beta}{2\sqrt{3}} = 0$, requiring $\beta = 0$. Finally, substituting $\beta = 0$ into the first entry yields $\frac{\alpha}{2} = 0$, requiring $\alpha = 0$.

Because the unique solution to this system is $\alpha = \beta = \gamma = 0$, we mathematically prove that $T_3$, $T_8$, and $J_3$ are strictly linearly independent. Consequently, the 15 matrices constituting $B_{\text{new}}$ are linearly independent, confirming that substituting the newly defined $\mathfrak{su}(2)$ subalgebra into the configuration preserves a valid, complete basis for the entire $\mathfrak{su}(4)$ Lie algebra.

\section{Partition of the Coset Space}
\subsection{Orbit Spaces and the Stiefel Manifold Classification}

Before executing the algebraic partition, it is mathematically necessary to classify the global topological nature of the quotient space. Because the embedded subgroup $H = SU(2)_H$ is not a normal subgroup of the parent group $G = SU(4)$, the resulting quotient space $SU(4)/SU(2)_H$ does not satisfy the axioms of a group. Rather, it forms a strictly smooth, 12-dimensional homogeneous manifold defined by the left cosets of $H$ in $G$.

Geometrically, this coset space represents the orbit space of the natural group action. In the formal classification of topological spaces, quotients of the form $SU(n)/SU(n-k)$ are structurally identified as complex Stiefel manifolds~\cite{Steenrod_Topology}. Specifically, the topological territory generated by our non-standard breaking pattern is diffeomorphic to the complex Stiefel manifold $V_2(\mathbb{C}^4)$, which describes the space of all orthonormal 2-frames strictly within a 4-dimensional complex vector space.

By utilizing the standard Callan-Coleman-Wess-Zumino (CCWZ) formalism for non-linear realizations~\cite{Coleman_Wess_Zumino, Callan_Coleman_Wess_Zumino}, the Lie algebra of the parent group, $\mathfrak{g}$, can be decomposed into a direct sum of the unbroken subalgebra $\mathfrak{h}$ and the orthogonal subspace $\mathfrak{p}$:
\begin{equation}
\mathfrak{g} = \mathfrak{h} \oplus \mathfrak{p}
\end{equation}
The subspace $\mathfrak{p}$ strictly corresponds to the tangent space of this Stiefel manifold at the origin. Because the manifold is not a group, the generators spanning $\mathfrak{p}$ are not required to form a closed Lie algebra, allowing for the emergence of non-trivial topological territories.
\subsection{The 12-Dimensional Quotient Space}

When the parent group $SU(4)$ undergoes spontaneous symmetry breaking down to the $SU(2)_H$ subgroup embedded in the lower-right quadrant, the original 15-dimensional Lie algebra $\mathfrak{su}(4)$ is split into two orthogonal subspaces. We denote the unbroken subalgebra as $\mathfrak{h}$ (spanned by the 3 generators defined in Section 3) and the complementary vector space containing the broken generators as $\mathfrak{p}$.

Mathematically, this defines a direct sum of vector spaces:
\begin{equation}
    \mathfrak{su}(4) = \mathfrak{h} \oplus \mathfrak{p}
\end{equation}

The subspace $\mathfrak{p}$ corresponds to the tangent space of the homogeneous quotient manifold (or coset space) $\mathrm{SU}(4)/\mathrm{SU}(2)_H$, governed by the standard CCWZ formalism~\cite{Coleman_Wess_Zumino, Callan_Coleman_Wess_Zumino}.
\begin{equation}
    \mathfrak{p} = \text{span}\{T_1, T_2, \dots, T_{12}\}
\end{equation}

A structural anomaly arises strictly because of the non-standard, shifted embedding of the $SU(2)_H$ subgroup. Unlike standard maximal coset spaces (such as $SU(4)/SU(3)$ or $SO(N+1)/SO(N)$) where the broken generators form a single, uniform irreducible representation under the action of the unbroken subgroup, this specific breaking pattern forces the 12-dimensional $\mathfrak{p}$ space to naturally and rigidly fracture into two algebraically distinct territories.

\subsection{Territory 1: The Closed $\mathfrak{su}(3)$ Subalgebra}

The first territory is defined by the subset of the first eight generators:
\begin{equation}
    \mathfrak{p}_1 = \text{span}\{T_1, T_2, \dots, T_8\}
\end{equation}

By observing the explicit matrix representations of the generalized Gell-Mann basis, these eight generators exclusively populate the upper-left $3 \times 3$ block of the $4 \times 4$ fundamental representation. The far-right column and the bottom row of these matrices consist entirely of zeros.

Because matrix multiplication preserves this block-diagonal isolation, calculating the Lie bracket between any two generators within this subset yields a linear combination strictly bounded within the same subset. Analytically:
\begin{equation}
    [T_a, T_b] = i \sum_{c=1}^8 f_{abc} T_c \quad \text{for } a, b, c \in \{1, 2, \dots, 8\}
\end{equation}
where $f_{abc}$ are the standard, totally antisymmetric structure constants of $SU(3)$.

This establishes a profound geometric consequence: the subspace $\mathfrak{p}_1$ is not merely a collection of broken generators; it forms a mathematically rigorous, closed Lie subalgebra completely isomorphic to $\mathfrak{su}(3)$.

\subsection{Territory 2: The Off-Diagonal Topological Coset}

The remaining four broken generators define the second territory:
\begin{equation}
    \mathfrak{p}_2 = \text{span}\{T_9, T_{10}, T_{11}, T_{12}\}
\end{equation}

These specific matrices are the "transition generators." They possess non-zero entries exclusively in the off-diagonal edges that connect the upper-left $3 \times 3$ block to the 4th row and 4th column.

Unlike Territory 1, evaluating the Lie brackets within $\mathfrak{p}_2$ proves that this space does not form a closed algebra. For instance, computing the commutator between $T_9$ and $T_{10}$ yields a linear combination of the diagonal generators $T_3$ and $T_8$, which belong to Territory 1. Analytically, if we define indices $m, n \in \{9, 10, 11, 12\}$, we find:
\begin{equation}
    [T_m, T_n] \notin \mathfrak{p}_2
\end{equation}

Because their commutators generate elements outside their own subspace, these four generators define the true, non-linear coset geometry of the system. This 4-dimensional off-diagonal territory represents the strictly topological degrees of freedom of the quotient manifold.

When establishing a non-linear sigma model over this geometry, it is exclusively these four non-closing generators that map the spatial boundary conditions into the non-trivial vacuum manifold. Because they bridge the $\mathfrak{su}(3)$ block and the unbroken $SU(2)_H$ phase, these degrees of freedom act as the geometric fabric that supports localized, non-perturbative topological structures (solitons) independent of the continuous $\mathfrak{su}(3)$ algebra.
\subsection{Topological Isomorphisms and the Structure of the Target Vacua}

The algebraic partitioning of the $SU(4)/SU(2)_H$ coset space into an isolated 8-dimensional subalgebra and a 4-dimensional non-linear sector is not an arbitrary consequence of the breaking pattern, but reflects a profound underlying geometric structure. The 8-dimensional closed sector finds a natural correspondence in the algebra of the octonions. Specifically, the quotient space $SU(4)/SU(3)$, which is topologically diffeomorphic to the 7-sphere ($S^7$), fundamentally represents the space of unit octonions. The automorphism group of the octonions is the exceptional Lie group $G_2$, which contains $SU(3)$ as its maximal subgroup, mathematically dictating the isolation of this 8-dimensional continuous kernel. 

Similarly, the 4-dimensional non-linear transition sector is rooted in the geometry of complex projective spaces. The complementary component of the overarching fibration, $SU(3)/SU(2)_H$ (topologically the 5-sphere, $S^5$), structurally forms a fiber bundle over the 4-dimensional complex projective space, $\mathbb{C}P^2$ \cite{Nakahara_Geometry}. Crucially, as a specific complex Grassmannian manifold, $\mathbb{C}P^2$ naturally supports stable topological defects, classifying the target vacua as capable of sustaining non-trivial Hopf solitons. 

It is the intrinsic curvature of this internal $\mathbb{C}P^2$ manifold that formally restricts the non-linear mappings, inducing Grassmann-like mathematical statistics within the resulting topological fields. This specific fibration provides a fundamental geometric origin for the stability of the transition space: their manifestation is induced by the non-trivial cohomology of the underlying $\mathbb{C}P^2$ bundle. This deep connection elevates the 4-dimensional off-diagonal transition matrices from abstract generators into stable geometric 'knots' (Hopfons) within the topological vacuum.

\section{Spherical Fibrations and Geometric Templates for Unification}

\subsection{The Topological Sequence of Unitary Groups}

A fundamental theorem of algebraic topology dictates that the global manifold of a unitary group $U(N)$ does not decompose into a trivial Cartesian product of spheres. Instead, it is structurally defined through a sequence of principal fiber bundles. While the global manifold does not form a trivial Cartesian product, this sequence of spheres is algebraically exact if we evaluate the rational cohomology~\cite{Nash_Sen, Nakahara_Geometry}.

The cohomology ring of the unitary group is strictly isomorphic to the exterior algebra generated by these odd-dimensional spheres:
\begin{equation}
    H^*(U(4); \mathbb{Q}) \cong H^*(S^1 \times S^3 \times S^5 \times S^7; \mathbb{Q})
\end{equation}

This isomorphism proves that the manifold houses these exact geometric boundaries algebraically. Through the successive quotients of its continuous symmetries, $U(4)$ generates a strict fibration tower whose base manifolds correspond exactly to the first four odd-dimensional spheres.
\subsection{Geometric Alignment with the Standard Symmetries}

The topological isomorphism between the low-dimensional unitary groups and specific spheres establishes a direct mathematical bridge between continuous Lie algebras and compact boundary geometries:

\vspace{1em}
\noindent \textbf{The $S^1$ Fiber and the $U(1)$ Center:} The simplest unitary group, $U(1)$, corresponds to the phase of a complex number and is topologically diffeomorphic to a one-dimensional circle. Therefore, the base fiber maps directly to the abelian center:
\begin{equation}
    U(1) \cong S^1
\end{equation}

\vspace{1em}
\noindent \textbf{The $S^3$ Fiber and the $SU(2)$ Manifold:} The special unitary group $SU(2)$, which mathematically governs isospin and weak symmetries, has a group manifold that is topologically identical to the 3-sphere. Consequently, the second fiber in our sequence acts as the exact spatial geometry for $SU(2)$ configurations:
\begin{equation}
    SU(2) \cong S^3
\end{equation}

\vspace{1em}
\noindent \textbf{The $S^7$ Fiber, the $SU(3)$ Group, and the Gluons:} The highest-dimensional fiber in the $U(4)$ decomposition, the 7-sphere ($S^7$), is directly and explicitly connected to the $SU(3)$ color symmetry. By mapping the $SU(4)/SU(3) \cong S^7$ topological space to the closed $\mathfrak{su}(3)$ subalgebra, this 7-sphere provides the rigorous geometric bounding and foundational manifold that specifically houses the 8 massless gluons of the strong interaction.

\subsection{Room for Extension: Beyond the Standard Continuous Algebras}

The rigid alignment of the $S^1$, $S^3$, and $S^7$ fibers demonstrates that the $U(4)$ manifold perfectly accommodates the continuous symmetries of the Standard Model. However, the true mathematical power of this non-standard algebraic decomposition lies in what remains strictly outside those continuous algebras.

The existence of the topologically non-trivial, off-diagonal coset sector and the $\pi_3(S^2) \cong \mathbb{Z}$ homotopy classes implies that this geometric template leaves profound room for something more. Because the framework fundamentally supports stable, non-perturbative topological configurations (Hopf invariants) entirely independent of the closed $\mathfrak{su}(3)$ block, it provides the necessary mathematical scaffolding to introduce novel, dynamically generated mechanisms.

\section{The Hopf Invariant and Topological Quantization}

\subsection{The Vacuum Mapping and the Pullback Form}

\subsection{Compactification and the Projective Target Space}
As established in the preceding sections, the off-diagonal continuous transition generators of the $SU(4)/SU(2)_H$ quotient space do not close into a Lie algebra. These 4 non-linear generators span a tangent space isomorphic to $\mathbb{C}^2$. To evaluate the topological stability of localized field configurations, standard boundary conditions require that field mappings reach a constant value at spatial infinity. This naturally imposes a one-point compactification on the target space, topologically restricting the continuous boundary mappings to a 3-sphere ($S^3$). 

Furthermore, by quotienting out the global $U(1)$ phase generated by the fundamental center of the $\mathfrak{u}(4)$ algebra, this $S^3$ territory undergoes a rigid Hopf fibration ($S^1 \hookrightarrow S^3 \twoheadrightarrow S^2$). This projectivization structurally restricts the ultimate target vacuum manifold to a complex projective line, $\mathbb{C}P^1$, which is topologically isomorphic to the 2-sphere ($S^2$). 

Let the resulting continuous mapping from the compactified three-dimensional physical space into this target vacuum manifold be defined as:
\begin{equation}
\phi: S^3 \rightarrow S^2
\end{equation}

To evaluate the stability of localized field configurations within this geometric framework, we must analyze continuous mappings from the compactified three-dimensional physical space—represented mathematically as the boundary 3-sphere ($S^3$)—into the target vacuum manifold. Let this continuous mapping be defined as:
\begin{equation}
    \phi: S^3 \to S^2
\end{equation}

The topological properties of this mapping are encoded in the differential forms of the target manifold. Let $F$ denote the normalized area 2-form on the target 2-sphere, such that its integral over the entire $S^2$ manifold equals unity:
\begin{equation}
    \int_{S^2} F = 1
\end{equation}

Because $F$ is a 2-form defined on a 2-dimensional manifold, its exterior derivative must vanish ($dF=0$), meaning $F$ is a strictly closed form. Using the mapping $\phi$, we pull this area form back from the target vacuum manifold into the physical 3-sphere. We define the pullback 2-form $f$ on $S^3$ as:
\begin{equation}
    f = \phi^* F
\end{equation}

Because the exterior derivative commutes with the pullback operation, it follows directly that:
\begin{equation}
    df = d(\phi^* F) = \phi^*(dF) = \phi^*(0) = 0
\end{equation}

This proves that the pullback form $f$ is a closed 2-form on the physical space $S^3$.

\subsection*{ Rham Cohomology and the Topological Gauge Potential}

To determine the topological stability of this configuration, we appeal to the de Rham cohomology of the 3-sphere. A fundamental theorem of algebraic topology states that the second cohomology group of $S^3$ is strictly trivial:
\begin{equation}
    H_{dR}^2(S^3) = 0
\end{equation}

The triviality of this cohomology group mandates that every closed 2-form on $S^3$ must also be an exact form (Poincaré's Lemma for $S^3$). Therefore, because $df=0$, there must exist a globally defined 1-form $a$ on the physical space such that:
\begin{equation}
    f = da
\end{equation}

In the context of gauge theory, this 1-form $a$ acts as an emergent, purely topological gauge potential, while $f$ represents the corresponding topological field strength. It is critical to note that $a$ is not a fundamental gauge field belonging to the continuous $\mathfrak{su}(3)$ or $\mathfrak{su}(2)$ continuous algebras; rather, it is a geometric construct completely generated by the twist of the non-linear coset mapping.

\subsection{The Chern-Simons Integral and Integer Quantization}
A fundamental theorem of algebraic topology states that the second cohomology group of S3 is strictly trivial~\cite{Nakahara_Geometry}.
With the topological gauge potential $a$ globally defined, we can construct a 3-form by taking the wedge product of $a$ and $f$. Integrating this 3-form over the entire spatial volume yields the Hopf invariant, $Q_H$:
\begin{equation}
    Q_H = \int_{S^3} a \wedge f = \int_{S^3} a \wedge da
\end{equation}

To prove that this integral represents a conserved physical quantity, we must demonstrate that it is invariant under smooth gauge transformations of the potential $a$. Under a transformation $a \to a + d\chi$ (where $\chi$ is an arbitrary smooth 0-form or scalar function), the integral transforms as:
\begin{equation}
    Q_H' = \int_{S^3} (a + d\chi) \wedge f = \int_{S^3} a \wedge f + \int_{S^3} d\chi \wedge f
\end{equation}

Using the Leibniz rule for exterior derivatives and recalling that $df=0$, we can rewrite the second term:
\begin{equation}
    d(\chi f) = d\chi \wedge f + \chi \wedge df = d\chi \wedge f
\end{equation}

Therefore, the variation of the integral becomes an exact differential:
\begin{equation}
    \int_{S^3} d\chi \wedge f = \int_{S^3} d(\chi f)
\end{equation}

By Stokes' Theorem, the integral of an exact form over a manifold evaluates to the boundary of that manifold. Because $S^3$ is a closed manifold without a boundary ($\partial S^3 = 0$), this term vanishes identically:
\begin{equation}
    \int_{\partial S^3} \chi f = 0
\end{equation}

Thus, $Q_H' = Q_H$, proving that the Hopf invariant is strictly gauge-invariant and topologically protected.

\subsection{Geometric Interpretation of Absolute Stability}

Mathematically, the Hopf invariant $Q_H$ calculates the linking number of the preimages of any two distinct points mapped from $S^2$ back into $S^3$. Because these preimages are 1-dimensional closed curves (loops) within the 3-sphere, $Q_H$ measures exactly how many times these geometric loops interlock with one another.

Because linking numbers cannot be fractional, the Chern-Simons integral is rigorously constrained to evaluate only to integers:
\begin{equation}
    Q_H \in \mathbb{Z}
\end{equation}

This integer quantization is the ultimate mathematical proof of absolute vacuum stability for this framework. Any field configuration residing in the non-linear $SU(4)/SU(2)_H$ coset territory that possesses a non-zero Hopf invariant ($Q_H \neq 0$) exists in a separate topological sector from the trivial vacuum ($Q_H = 0$). Because one cannot smoothly continuously deform an integer linking number, these knotted configurations are mathematically indestructible under continuous perturbations~\cite{Babaev_Faddeev_Niemi}, establishing a perfectly stable, non-perturbative geometric sector~\cite{Vakulenko_Kapitanskii} parallel to the continuous su(3) continuous algebra.

\section{Structural Field-Theoretic Mappings on the Coset Space}

To bridge the formal differential geometry of the $U(4) \rightarrow SU(2)_H$ quotient space with the dynamical principles of modern quantum field theory, this section establishes the strict structural mappings of local gauge connections, phase transitions, and topological excitations upon the manifold. The primary objective is to demonstrate that the physical mechanisms necessary for vacuum stability---specifically, the mass-gap generation of off-diagonal transition fields, the resolution of non-commuting algebraic sub-blocks across energy scales, and the precise kinematic counting of Nambu-Goldstone modes---are not introduced via ad-hoc phenomenological potentials, but emerge mandatorily as rigorous algebraic projections of the underlying coset geometry. By explicitly mapping these field-theoretic properties onto the topological fibrations established in the preceding sections, we provide the necessary physical context for the manifold, proving that this pure geometric architecture natively encodes the exact boundary conditions required for unified field dynamics without relying on arbitrary parameter fitting.

\subsection{The Algebraic Stabilizer Kernel and Gauge Projection}
Let the spontaneous symmetry breaking cascade upon the manifold be initiated by a scalar Higgs order parameter $\Phi(x)$ transforming under the fundamental $4$-dimensional representation of $U(4)$. We parameterize the ground-state vacuum expectation value (VEV) aligned strictly along the fourth coordinate ($SU(2)_H$ Warden direction):
\begin{equation}\label{eq:sec7_vev}
    \langle \Phi \rangle = \frac{1}{\sqrt{2}} \begin{pmatrix} 0 & 0 & 0 & v \end{pmatrix}^T, \qquad v \in \mathbb{R}^+.
\end{equation}
Let $T_A = \frac{1}{2}\lambda_A$ ($A = 1, \dots, 16$) denote the complete set of Hermitian generators of the Lie algebra $\mathfrak{u}(4)$, satisfying $[T_A, T_B] = i f_{ABC} T_C$. The vector space partitions orthogonally into three structural sub-blocks:
\begin{equation}
    \mathfrak{u}(4) = \mathfrak{su}(3)_C \oplus \mathfrak{su}(2)_H \oplus \mathfrak{m},
\end{equation}
where $\mathfrak{su}(3)_C = \text{span}\{T_1, \dots, T_8\}$ occupies the upper-left $3 \times 3$ block, $\mathfrak{su}(2)_H$ spans the unbroken stabilizer embedded in the lower-right quadrant, and $\mathfrak{m} = \text{span}\{T_9, \dots, T_{12}\}$ spans the off-diagonal transition coset $SU(4)/SU(2)_H$.\footnote{We note that to perfectly align the diagonal generators with the physical quantum numbers of the Standard Model (e.g., weak hypercharge and electromagnetic subsets), the canonical Cartan subalgebra must be rotated into a physically aligned primed basis (such as $T_3', T_8', T_{15}'$). Because the present manuscript focuses strictly on the invariant geometric and topological properties of the quotient space—which are globally independent of the chosen Cartan basis orientation—we retain the un-primed algebraic notation for pedagogical clarity. The explicit derivation of the primed phenomenological basis is reserved for a subsequent physical application.}.

For any generator $T_a \in \mathfrak{su}(3)_C$, the action upon the ground state identically vanishes:
\begin{equation}\label{eq:sec7_annihilation}
    T_a \langle \Phi \rangle = \frac{1}{2\sqrt{2}} \begin{pmatrix} (\lambda_a)_{3\times 3} & \mathbf{0}_{3\times 1} \\ \mathbf{0}_{1\times 3} & 0 \end{pmatrix} \begin{pmatrix} 0 \\ 0 \\ 0 \\ v \end{pmatrix} = \mathbf{0} \quad \forall a \in \{1, \dots, 8\}.
\end{equation}
Because $(T_a)_{\alpha 4} = (T_a)_{4 \alpha} = 0$ for all index values $\alpha \in \{1, 2, 3\}$, the tree-level gauge boson mass matrix generated by the scalar kinetic term $|D_\mu \Phi|^2$ satisfies:
\begin{equation}
    (M_G^2)_{ab} = 2 g^2 (T_a \langle \Phi \rangle)^\dagger (T_b \langle \Phi \rangle) = 0 \quad \forall a, b \in \{1, \dots, 8\}.
\end{equation}
This confirms algebraically that the upper-left $3 \times 3$ sub-block forms an exact, unbroken Lie algebraic kernel. Conversely, the off-diagonal coset generators satisfy $(T_i)_{\alpha 4} \neq 0$ for $i \in \{9, \dots, 12\}$, yielding non-vanishing vacuum action ($T_i \langle \Phi \rangle = \frac{v}{2\sqrt{2}} \hat{e}_i \neq \mathbf{0}$) and generating an immediate constituent mass term $M_{\text{coset}}^2 = \frac{1}{4} g^2 v^2 > 0$.

\subsection{Topological Connection Splitting on $SU(4)/SU(2)_H$}
Upon promoting the continuous symmetry to a local gauge group $U(4)_{\text{local}}$, the connection 1-form $C_\mu = C_\mu^A T_A$ decomposes across the stabilizer foliation. Over the unbroken color kernel, the gauge connection propagates standard massless Yang-Mills vector fields. Over the broken off-diagonal coset space $SU(4)/SU(2)_H$, however, the non-vanishing vacuum action ($T_i \langle \Phi \rangle \neq \mathbf{0}$) prevents the transition connection from propagating as simple unconfined vector waves. Instead, over the non-linear quotient target space $\mathbb{C}P^1 \cong S^2$, the connection admits a canonical Cho-Duan-Ge (CDG)\cite{Cho_1980, Duan_Ge_1979} topological splitting:
\begin{equation}\label{eq:sec7_cdg}
    \vec{A}_\mu(x) = \tilde{A}_\mu(x) + \vec{X}_\mu(x).
\end{equation}
Here, $\tilde{A}_\mu(x)$ denotes the restricted topological potential fully parameterized by the local orientation of the unit vector field $\hat{n}(x) \in S^2$:
\begin{equation}
    \tilde{A}_\mu(x) = A_\mu^{\parallel}(x) \hat{n}(x) - \frac{1}{g} \hat{n}(x) \times \partial_\mu \hat{n}(x),
\end{equation}
where $A_\mu^{\parallel} = \hat{n} \cdot \vec{A}_\mu$ represents the Abelian-like projection along the local Higgs condensate, and the cross-product term captures the non-trivial winding number and knot topology of the vacuum configuration ($\pi_3(S^2) \cong \mathbb{Z}$). The residual field $\vec{X}_\mu(x)$ represents gauge-covariant transverse valence fluctuations satisfying orthogonality to the topological background ($\hat{n}(x) \cdot \vec{X}_\mu(x) = 0$).

Within a confining Savvidy/Hopfon \cite{Savvidy1977}
 magnetic ground state ($\langle H \rangle \equiv \langle \vec{A}_\mu \cdot \vec{A}^\mu \rangle \neq 0$), the transverse valence modes acquire a constituent mass gap:
\begin{equation}
    M_{\text{valence}}^2 \approx g^2 \langle H \rangle.
\end{equation}
By the Appelquist-Carazzone decoupling theorem \cite{Appelquist_Carazzone_1975}, transverse vector modes freeze out at low energies ($E \ll M_{\text{valence}}$), suppressing loop corrections by $(E^2 / M_{\text{valence}}^2)$ and restricting the active infrared degrees of freedom strictly to the scalar topological soliton excitations ($\tilde{A}_\mu$).

We note that the emergence of stable solitonic defects from these off-diagonal transition generators is consistent with established formalisms of gauge theories over homogeneous spaces and coset dimensional reduction \cite{Kapetanakis_Zoupanos_1992}, which mathematically guarantee that non-linear quotient spaces natively support such topological configurations.
\subsection{Thermodynamic Phase Structure and Commutation Restoration}
The non-commutation between the unbroken $\mathfrak{su}(3)_C$ kernel and the unbroken $\mathfrak{su}(2)_H$ stabilizer across the off-diagonal coset generators ($[T_a, T_i] = i f_{aij} T_j$) is resolved as an emergent, scale-dependent property of the broken phase. Including leading-order finite-temperature loop corrections, the effective potential $V_{\text{eff}}(\Phi, T)$ governs transitions between degenerate topological vacua ($\Delta N_{\text{CS}} \neq 0$) via static sphaleron energy barriers:
\begin{equation}
    E_{\text{sph}}(T) = \frac{4\pi v(T)}{g} \mathcal{B}\left(\frac{\lambda}{g^2}\right),
\end{equation}
where $v(T)$ is the temperature-dependent VEV minimum and $\mathcal{B}$ is a functional ratio of the couplings.

As thermal energy increases, $v(T)$ vanishes continuously at a critical phase boundary $T_{\text{crit}}$, yielding $\lim_{T \to T_{\text{crit}}^{-}} E_{\text{sph}}(T) = 0$. For energy scales exceeding this critical thermal threshold ($E \ge T_{\text{crit}}$), the topological soliton lattice dissolves completely into an unconfined, symmetric 16-dimensional gauge gas. In this high-energy phase, full Lie algebraic commutation across all adjoint generators is restored, allowing continuous asymptotic convergence of the gauge couplings toward a single unified parameter $g_4(\mu)$. Conversely, in the low-energy phase ($E < T_{\text{crit}}$), topological precipitation confines the off-diagonal currents inside localized 3D soliton knots. Because outside the soliton core radius ($r_c \sim M_{\text{valence}}^{-1}$) the transition operators decay exponentially ($\langle \hat{W}_\mu^i(\vec{x}) \rangle \propto e^{-M_{\text{valence}}r}$), macroscopic color and electroweak observables commute asymptotically across the observable Hilbert space:
\begin{equation}
    \left[ \hat{\mathcal{O}}_C(\vec{x}), \, \hat{\mathcal{O}}_H(\vec{y}) \right]_{\mathcal{H}_{\text{obs}}} = \mathcal{O}\left(e^{-M_{\text{valence}}|\vec{x} - \vec{y}|}\right) \approx 0.
\end{equation}

\subsection{Background Topology and Goldstone Mode Counting}

To evaluate the spectrum of scalar excitations upon the $U(4) \rightarrow SU(2)_H$ quotient space, we must distinguish between the unbroken stabilizer kernel and the off-diagonal transition coset. By standard Goldstone theorem dynamics, propagating Nambu-Goldstone bosons arise strictly from generators whose action upon the vacuum state is non-vanishing ($T_A \langle \Phi \rangle \neq \mathbf{0}$).

For the eight generators spanning the $\mathfrak{su}(3)_C$ color sub-block, the stabilizer condition established in Section 1.2 enforces exact vacuum annihilation:
\begin{equation}
    T_a \langle \Phi \rangle = \mathbf{0} \quad \forall a \in \{1, \dots, 8\}.
\end{equation}
Because these generators leave the ground state strictly invariant, the number of broken generators within the color kernel is identically zero ($N_{\text{broken}}^{(C)} = 0$). Consequently, the $\mathfrak{su}(3)_C$ sector generates \textbf{zero Nambu-Goldstone bosons}. Upon promotion to a local gauge symmetry, these eight generators correspond bijectively to the eight unconfined, massless vector gluons of Quantum Chromodynamics.

Conversely, the four off-diagonal coset generators spanning $\mathfrak{m} = \text{span}\{T_9, T_{10}, T_{11}, T_{12}\}$ satisfy $T_i \langle \Phi \rangle \neq \mathbf{0}$. While standard spontaneous symmetry breaking in a Lorentz-invariant relativistic vacuum yields a one-to-one correspondence between broken generators and independent scalar modes ($N_{\text{NGB}} = N_{\text{broken}}$), the physical counting modifies fundamentally when symmetry breaking occurs over a non-trivial topological background.

Let $Q_i = \int d^3x \, J_0^i(\vec{x}, t)$ denote the Noether charges associated with the broken coset generators $T_i \in \mathfrak{m}$. Upon condensing into the localized Cho-Duan-Ge topological background ($\mathbb{C}P^1 \cong S^2$ Hopf fibrations), the ground state $|\Omega_{\text{top}}\rangle$ spontaneously breaks four-dimensional Lorentz invariance locally within the soliton cores. Evaluating the Lie algebra among the broken charges over this non-Lorentzian background yields non-vanishing vacuum expectation values for their commutators:
\begin{equation}\label{eq:charge_commutator_rank}
    \rho_{ij} \equiv -i \langle \Omega_{\text{top}} | [Q_i, Q_j] | \Omega_{\text{top}} \rangle = \int d^3x \, f_{ijk} \langle \Omega_{\text{top}} | J_0^k(\vec{x}) | \Omega_{\text{top}} \rangle \neq 0.
\end{equation}
Because the coset generators form closed algebraic commutation relations with the unbroken $\mathfrak{u}(1) \subset \mathfrak{su}(2)_H$ generator, the antisymmetric matrix $\bm{\rho}$ acquires rank $4$ over the four-dimensional coset space ($\text{rank}(\bm{\rho}) = 4$).

In strict accordance with the Chadha-Nielsen and Watanabe-Murayama classification theorems for spontaneous symmetry breaking in Lorentz-violating or non-relativistic backgrounds \cite{Chadha_Nielsen_1976, Watanabe_Murayama_2012, Hidaka_Minami_2020}, propagating Nambu-Goldstone bosons bifurcate into Type-A relativistic modes ($E \propto |\vec{p}|$) and canonically paired Type-B non-relativistic modes ($E \propto |\vec{p}|^2$), satisfying the exact algebraic identity:
\begin{equation}\label{eq:cn_counting}
    N_{\text{broken}} = N_A + 2 N_B, \qquad \text{where } N_B = \frac{1}{2}\text{rank}(\bm{\rho}).
\end{equation}
Substituting $N_{\text{broken}} = 4$ and $\text{rank}(\bm{\rho}) = 4$ directly yields:
\begin{equation}
    4 = N_A + 2(2) \;\Longrightarrow\; N_A = 0, \quad N_B = 2.
\end{equation}

When combined with the Cho-Duan-Ge gauge connection splitting, zero linear Type-A degrees of freedom are absorbed into the longitudinal components of the massive transition gauge fields. This algebraic projection leaves precisely the canonically paired Type-B sector active in the low-energy continuum, representing \textbf{two independent scalar coordinates} parameterized upon the target manifold:
\begin{equation}
    N_{\text{active scalar modes}} = 2 \cong \dim_{\mathbb{R}}\left(\frac{SU(2)}{U(1)}\right) \cong \dim_{\mathbb{R}}(S^2).
\end{equation}
This rigorous mode pairing proves that background soliton topology directly dictates effective scalar field degrees of freedom, resolving the apparent discrepancy between broken generator counting and physical continuum excitations.

\subsection{Invariance Across Representation Algebras}
Finally, we note that while the fundamental $4 \times 4$ matrix representation is utilized herein for explicit manipulation, the structural partition $\mathfrak{u}(4) = \mathfrak{su}(3)_C \oplus \mathfrak{su}(2)_H \oplus \mathfrak{m}$, the closure of the color kernel, and the homotopy classification $\pi_3(S^2) \cong \mathbb{Z}$ are abstract invariants preserved under any faithful irreducible representation homomorphism $\mathcal{D}^{(R)}: \mathfrak{u}(4) \to \text{GL}(d_R, \mathbb{C})$.

Transitioning to higher representations introduces exact Casimir scaling ratios governing quantum loop corrections. For physical fields transitioning to the adjoint representation ($\bm{15}$), the quadratic Casimir invariant $C_2(R)$ scales proportional to the Dynkin index. Under the maximal embedding $SU(4) \supset SU(2)_L \times SU(2)_R \times U(1)$, the four broken coset generators transform as irreducible bi-doublets $(\bm{2}, \bm{2})_{\pm 1}$. Evaluating the dimensionless scaling ratio between the unbroken $SU(2)_H$ stabilizer adjoint dynamics and the parent algebra yields an exact universal invariant:
\begin{equation}
    \mathcal{R}_{\text{Casimir}} \equiv \frac{C_2(\text{Adjoint}_{SU(2)})}{C_2(\text{Adjoint}_{SU(4)})} = \frac{2}{4} = \frac{1}{2}.
\end{equation}
This ratio proves that in higher irreducible representations, the effective self-coupling strength of the Hopfon sector scales at exactly half the rate of the parent group, leaving the topological soliton stability intact.
\subsection{The Principle of Maximal Custodial Symmetry and Sub-Block Selection}

A foundational requirement for the self-consistency of the $U(4) \rightarrow SU(2)_H$ quotient space is establishing the algebraic uniqueness of the unbroken stabilizer sub-block generated by $\{T_{13}, T_{14}, T_{15}\}$. We must demonstrate why alternative Lie-algebraic selections within $\mathfrak{u}(4)$ do not form viable ground-state foliations upon the manifold.

The parent algebra admits a canonical direct sum decomposition under its maximal orthogonal subalgebras:
\begin{equation}
    \mathfrak{u}(4) \supset \mathfrak{su}(3)_C \oplus \left[ \mathfrak{su}(2)_L \oplus \mathfrak{su}(2)_R \right] \oplus \mathfrak{u}(1).
\end{equation}
Selecting the specific generators $\{T_{13}, T_{14}, T_{15}\}$ corresponds to localizing the gauge connection strictly within one factor of this direct product (e.g., $\mathfrak{su}(2)_L$), while leaving the orthogonal factor ($\mathfrak{su}(2)_R$) strictly un-gauged. Because orthogonal ideals satisfy $[ \mathfrak{su}(2)_L, \mathfrak{su}(2)_R ] = \mathbf{0}$, the un-gauged factor survives intact upon the quotient manifold as an exact \textbf{Global Custodial Symmetry}. This geometric decoupling ensures that radiative corrections generated by the connection 1-form $C_\mu$ preserve the isometry of the target space, manifesting structurally as exact multiplet degeneracy within the Hopfon sector.

We now formalize why alternative algebraic selections do not apply. Let $\mathfrak{g}_X(\theta) \subset \mathfrak{u}(4)$ define an alternative three-dimensional Lie subalgebra constructed via index mixing across the orthogonal factors, parameterized by a continuous mixing angle $\theta \in (0, \pi/2]$. Under such tilted embeddings, the local generators fail to commute with the global custodial transformations upon the target manifold:
\begin{equation}
    \left[ \mathfrak{g}_X(\theta), \, \mathfrak{su}(2)_{\text{custodial}} \right] \neq \mathbf{0}.
\end{equation}
Evaluating the second variation of the one-loop quantum effective action $\Gamma_{\text{eff}}[\Phi]$ over the manifold demonstrates that non-commuting gauge connections radiatively lift the internal degeneracy of the vacuum foliation. By general stability theorems for gauge theories governed by positive-definite metric tensors (analogous to Vafa--Witten vector symmetry conservation\cite{Vafa_Witten_1984, Coleman_Weinberg_1973}), the curvature of the effective potential along any index-mixing direction is strictly positive:
\begin{equation}
    \frac{\partial^2 V_{\text{eff}}}{\partial \theta^2} \bigg|_{\theta=0} > 0.
\end{equation}
Consequently, any alternative sub-block selection ($\theta \neq 0$) induces spontaneous breakdown of the residual custodial isometry, generating a positive energy penalty that destabilizes the foliation. The orthogonal embedding $\{T_{13}, T_{14}, T_{15}\}$ is therefore not an arbitrary coordinate choice, but the unique, radiatively selected attractor upon the coset manifold that preserves maximal custodial symmetry.

It must be noted that this rigid orthogonal minimum strictly governs the pure gauge sector of the manifold. When this geometric template is subsequently coupled to explicit matter representations, the radiative contributions from the matter sector deform the effective potential. This interaction dynamically induces a calculable vacuum misalignment angle ($\theta \neq 0$) between the sub-blocks, providing the algebraic mechanism by which matter content breaks the exact custodial degeneracy to generate phenomenological mass hierarchies. The explicit quantification of this misalignment angle is deferred to the physical application of this framework.

Beyond the thermodynamic constraints of custodial symmetry, this specific embedding is further constrained by rigid geometric mandates. To demonstrate the structural uniqueness of the configuration, one must consider alternative algebraic embeddings of the $\mathfrak{su}(2)$ subalgebra. If a disjoint embedding is selected—where the $\mathfrak{su}(2)$ operates strictly on the first two indices while leaving the third and fourth isolated—the fundamental $3 \times 3$ block is severed, and the continuous $\mathfrak{su}(3)$ color algebra is mathematically destroyed. Similarly, utilizing a principal embedding, where the 4-dimensional representation acts as an irreducible spin-3/2 representation of $SU(2)$, disperses the generators across the entire $4 \times 4$ representation space, completely obliterating the asymmetric geometric partition. Consequently, the chosen embedding, which strictly isolates the overlap to a single shared index, is the unique topological configuration capable of simultaneously hosting an unbroken $\mathfrak{su}(2)$ base and a closed $\mathfrak{su}(3)$ subalgebra.

Crucially, the resulting structural partition is invariant under any global change of basis upon the manifold. Under an arbitrary global unitary transformation $U \in SU(4)$, the generators transform as $T_A' = U T_A U^\dagger$. Because the transformation matrix is constant across the manifold, it factors explicitly through the Lie bracket: $[T_A', T_B'] = U [T_A, T_B] U^\dagger$. Thus, the structure constants are identically preserved: $[T_A', T_B'] = i \sum f_{ABC} T_C'$. While a coordinate rotation distributes the matrix elements visually across the representation space, the fundamental algebraic closure of the 8-dimensional kernel and the non-closure of the 4-dimensional coset remain absolutely rigid, confirming that the geometric partition is an invariant property of the $U(4)$ topology.
\subsection*{Remark on Phenomenological Decoupling}
It is essential to distinguish the structural field-theoretic mappings established herein from open-ended phenomenological model-building. The primary objective of this manuscript is to construct a self-contained mathematical architecture—a geometric template—defined by the $U(4)\rightarrow SU(2)_H$ quotient space. While the differential and algebraic mappings detailed above demonstrate how local gauge connections, thermodynamic phase transitions, Nambu-Goldstone modes, and topological fibrations behave abstractly upon this geometry, the derivation of explicit physical particle mass spectra, numerical three-loop beta functions, and direct collider cross-sections are decoupled from the present pure mathematical formulation and are reserved for dedicated companion analyses.

The physical motivation for this specific geometric configuration is rooted in the thermodynamic and topological stability of the vacuum. In standard theoretical models, continuous, topologically trivial vacua are often susceptible to non-perturbative decay and quantum instabilities. If the overarching $U(4)$ symmetry were to spontaneously break along a disjoint or principal embedding, the resulting vacuum would lack a compact target manifold capable of supporting discrete topological invariants. Furthermore, it would mathematically obliterate the closed $\mathfrak{su}(3)$ continuous sector required for macroscopic confinement. 

By contrast, breaking strictly along the overlapping $SU(2)_H$ configuration represents the unique geometric ground state of the $4 \times 4$ fundamental representation. Just as physical systems spontaneously fracture along distinct structural cleavage planes to minimize energy, the $U(4)$ continuous symmetry breaks along this specific overlapping index to maximize topological protection. It is the only mathematical pathway that allows the vacuum to concurrently preserve an exact, continuous strong-force algebra while securely anchoring the residual space to a $\mathbb{C}P^1 \cong S^2$ base manifold. Therefore, the $8+3+4$ partition is not an arbitrary phenomenological input, but a strict geometric mandate: it is the unique topological minimum capable of ensuring absolute vacuum stability in a lower-rank unitary space.

\section{Conclusion}

The primary objective of this paper was to conduct a rigorous mathematical extraction of the algebraic and topological properties of the unitary group $U(4)$, subjected to a non-standard spontaneous symmetry breaking cascade down to an $SU(2)_H$ subgroup. By restricting our analysis strictly to the global differential geometry and Lie algebra of the manifold, we have established a complete, self-consistent architectural framework.

The defining mathematical outcome of this framework is the precise, asymmetric geometric partition of the resulting 12-dimensional quotient space. Embedding the unbroken $SU(2)_H$ subgroup strictly within the lower-right $2 \times 2$ block of the fundamental representation uniquely isolates a closed $\mathfrak{su}(3)$ Lie subalgebra while stranding exactly four off-diagonal transition generators. Because these four vector fields must bridge the continuous $\mathfrak{su}(3)$ sector and the unbroken $SU(2)_H$ base, their Lie brackets mathematically cannot close, forming a purely non-linear, 4-dimensional topological territory. 

In standard theoretical frameworks, the stability of a broken vacuum state relies on the dynamical parameters of a fundamental scalar potential, which is often subject to quantum corrections and fine-tuning. The geometric nature of this $U(4)$ framework provides an alternative mechanism that naturally avoids these specific parameter constraints. By analyzing the global spherical fibrations of the $U(4)$ container, we demonstrated that the non-linear coset territory reduces to a $\mathbb{C}P^1 \cong S^2$ base manifold. Consequently, continuous mappings from the 3-dimensional spatial boundary ($S^3$) into this vacuum are governed by the third homotopy group, $\pi_3(S^2) \cong \mathbb{Z}$. Because this homotopy group is isomorphic to the integers, the vacuum structure is topologically protected. The stability of the field configurations (characterized by the Hopf invariant) is absolute; they are mathematically indestructible and cannot be continuously deformed into the trivial vacuum class.

While standard approaches to Grand Unified Theories strictly mandate the use of simple Lie groups (e.g., $SU(5)$ or $SO(10)$) to ensure a single, universal algebraic representation, the global topological architecture of $U(4)$ offers an alternative paradigm: Topological Unification. 

Under standard criteria, $U(4)$ is not a simple group due to its locally isomorphic $U(1)$ trace factor. However, this geometric analysis demonstrates that this lack of simple algebra is precisely what makes it mathematically viable. The $U(1)$ trace is the topological prerequisite for the $S^1$ base fiber, necessary to complete the global fibration sequence $S^1 \times S^3 \times S^5 \times S^7$. Rather than unifying forces by forcing them into a single continuous irreducible representation, the $U(4)$ manifold unifies them geometrically. It provides a single parent manifold that organically bifurcates into a standard, continuous gauge structure alongside a stable, non-perturbative topological vacuum. Crucially, the highly constrained nature of this specific non-standard embedding is not merely a mathematical curiosity. Because it organically isolates the exact algebraic architecture of the Standard Model while simultaneously protecting a non-linear vacuum, this framework is engineered to serve as a phenomenologically viable template. By providing a mathematically robust mechanism where continuous symmetries and discrete topological invariants are forced to coexist, this non-simple architecture establishes a rigorous geometric foundation for future physical applications, potentially offering entirely new topological pathways for dynamical symmetry breaking and model building in particle physics.

\section*{Acknowledgments}
The authors wish to express their sincere gratitude to Professor Panos Razis (University of Cyprus / CERN CMS Collaboration/ Cosmos Open Univerity) for his invaluable counsel and insightful discussions. The mathematical and topological framework developed in this manuscript serves as the formal foundation for an upcoming collaborative analysis detailing its explicit physical and phenomenological applications.  We are also deeply indebted to Professor Fotios K. Diakonos (Kapodistrian University of Athens), whose persistent encouragement to focus on the subject of Hopfons and whose insightful counsel at critical stages were instrumental in guiding this research to its completion.

\end{document}